\newcommand{\xfe}{[$X$/Fe]}
\newcommand{\feih}{[Fe/H]$_{\rm I}$}
\newcommand{\feiih}{[Fe/H]$_{\rm II}$}
\def\aa{{A\&A}}
\def\aap{{A\&A}}
\def\aas{{A\&AS}}
\def\aj{{AJ}}
\def\annrev{{ARA\&A}}
\def\araa{{ARA\&A}}
\def\apj{{ApJ}}
\def\apjs{{ApJS}}
\def\baas{{BAAS}}
\def\mnras{{MNRAS}}
\def\nat{{Nature}}
\def\pasp{{PASP}}

\documentclass[cup5b]{caps}
\usepackage{graphicx}
\usepackage{amssymb}
\usepackage{ociwsymp4e}  
\HeadText{Ivans et al}

\begin{document}

\pagenumbering{arabic}


%
%
\author[]{I.\ I.\ IVANS$^{1}$, C.\ SNEDEN$^{2}$, C.\ R.\ JAMES$^{3}$, G.\ W.\ PRESTON$^{4}$, \\
J.\ P.\ FULBRIGHT$^{4}$, P.\ A.\ H\"OFLICH$^{2}$, B.\ W.\ CARNEY$^{5}$, and J.\ C.\ WHEELER$^{2}$ 
\\
(1) California Institute of Technology, Pasadena, CA, USA\\
(2) The University of Texas at Austin and McDonald Observatory, TX, USA\\
(3) Sam Houston State University, Huntsville, TX, USA\\
(4) The Observatories of the Carnegie Institution of Washington, Pasadena, CA, USA\\
(5) University of North Carolina, Chapel Hill, NC, USA}

\chapter{A New Population of Old Stars}

\begin{abstract}
We report the results of a coherent study of three chemically anomalous 
metal-poor ([Fe/H] $\sim$ --2) stars.  These objects exhibit unusually 
low abundances of Mg, Si, Ca ($\alpha$-elements) and Sr, Y, and Ba
(neutron-capture elements).  Our analyses confirm and expand upon 
earlier reports of atypical abundances in BD+80~245, G4-36, and 
CS22966-043.  We also find that the latter two stars exhibit enhanced 
abundances of Cr, Mn, Ni, and Zn (iron-peak elements), along with what 
appears to be large abundances of Ga, with respect to the abundance of 
iron.  Comparing the chemical abundances of these stars to supernova 
model yields, we derive supernovae ratios of Type~Ia versus Type~II 
events in the range of 0.6 $\le$ $(N_{Ia}/N_{II})_{NewPop}$ $\le$ 1.3.  
Whereas, for the Sun, we derive supernovae ratios in good agreement 
with those found in the literature: 0.18 $\pm$ 0.01 $<$ 
$(N_{Ia}/N_{II})_{\odot}$ $<$ 0.25 $\pm$ 0.06.  Given the relatively low 
metallicity and high $(N_{Ia}/N_{II})$ ratios of the low-$\alpha$ 
stars studied here, these objects may have witnessed, or been born 
from material produced in the yields of the earliest supernova Type~Ia 
events.
\end{abstract}

\section{Introduction \& Background\label{intro}}

At early Galactic times, the main contributors to the chemical 
enrichment of the interstellar medium were presumably the most massive 
and shortest-lived stars.  Their resulting core collapse supernovae 
explosions of Type~II (``SNe~II''; and to a lesser extent, SNe~Ib and
SNe~Ic), enriched the medium in $\alpha$-elements (eg.\ 0, Ne, Mg, Si, 
Ca) and odd-Z elements (eg.\ Na, Al, P), among others.  Neutron-rich 
sites, presumably associated with the core collapse supernovae,
contributed additional elements produced in rapid neutron-capture 
nucleosynthesis (``$r$-process''; eg.\ Eu).  Evolving more slowly, 
lower mass stars began contributing their ejecta to the medium at later 
Galactic times, first adding those isotopes produced in slow 
neutron-capture nucleosynthesis (``$s$-process''; eg.\ Sr, Ba) and then 
later, Fe-peak nuclei via supernovae Type~Ia (``SNe~Ia'') explosions. 

The stage in Galactic evolution at which SNe~Ia began to contribute is
not known precisely.  However, an overall Galactic trend in the lowering
of the [$\alpha$/Fe]-ratio at about [Fe/H] $\simeq$ --1 is associated
with the increase in Fe due to SNe~Ia contributions (eg.\ Tinsley 1979, 
Pagel \& Tautvaisiene 1995, their Figure~3).

A number of studies have discovered stars of intermediately low 
metallicities that also possess relatively low [$\alpha$/Fe], [Na/Fe],
and/or neutron-capture element abundance ratios when compared to stars 
of comparable metallicities (with differences in the range of 0.1 -- 
0.6~dex) .  The studies include the outer halo star BD+6~855 ([Fe/H] 
$\simeq$ --0.7; Carney \& Latham 1985); a sample of high velocity stars 
(--1.3 $<$ [Fe/H] $<$ --0.4; Nissen \& Schuster 1997; ``NS97''); the 
outer halo common proper motion pair HD134439/40 ([Fe/H] $\simeq$ 
--1.5; King 1997); and the relatively young outer halo clusters Pal~12 
and Rup~106 ([Fe/H] $\simeq$ --1 and --1.4, respectively; Brown et al 
1997).  The abundances and kinematics of all of these objects are 
consistent with the idea of accretion events in the outer halo.

Studies encompassing larger samples of halo stars have found that the
average [$\alpha$/Fe]-ratio is lower for stars with higher-than-average
space velocities (Fulbright 2002) and for stars with orbits that take 
them to the outer halo (Stephens \& Boesgaard 2002).  These 
kinematically extreme stars may have formed at large distances from the 
Galactic center in either localized star-forming regions or in 
proto-galactic fragments which were later assimilated into the Milky
Way.

Recently, there have been a few serendipitous discoveries of 
low-metallicity stars ([Fe/H] $\simeq$ --2) that exhibit anomalously
low abundances of $\alpha$-elements (in some cases, over 1 dex lower
than is exhibited in halo stars of comparable metallicities that 
follow the general trend of an overall $\alpha$-enhancement).  
BD+80~245 (Carney et al 1997), G4-36 (James 1998, 2000), and 
BPS~CS22966-043 (Preston \& Sneden 2000) all exhibit similar 
under-abundances in their [$\alpha$/Fe]-ratios and are also 
deficient in the abundance of neutron-capture elements.  In our
analysis, we studied these stars as a group, expanding the number of
elemental abundances, and explored the idea that the elemental 
abundance ratio diminishments in these stars are the results of iron 
enhancements from SNe~Ia contributions.

\section{Key Abundance Results}

\begin{figure}
  \centering
  \includegraphics[width=13.3cm,angle=0]{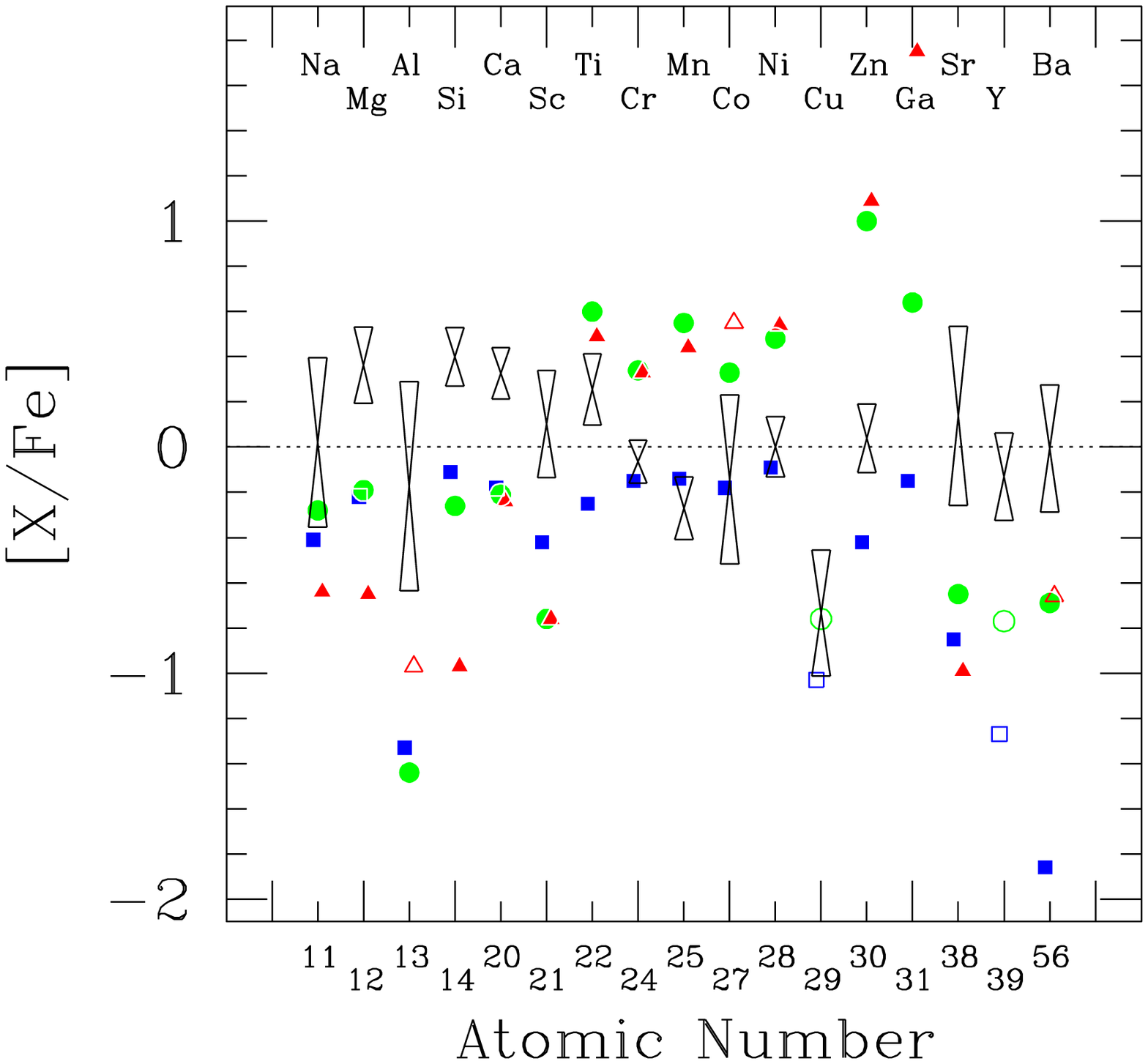}
    \caption{Elemental abundances from Table~\ref{table1} for BD+80~245 
(blue square), G4-36 (green circle) and CS22966-043 (green triangle).  
Abundances of halo stars of comparable metallicities (--1.75 $<$ [Fe/H] 
$<$ --2.25) are denoted by ($\bowtie$, centered on and extending 
1-$\sigma$ from the mean value).  Strict upper limits are denoted by
hollow symbols.}
    \label{figure1}
  \end{figure}

In our analysis, we confirm the previous reports of the unusual
nature of these stars, and expand the number of elements studied.
All three objects possess anomalously low light-$\alpha$- and 
odd-Z element abundances with respect to stars of comparable 
metallicities, as well as low abundances of Sr and Ba (by 
almost 2~dex in the case of BD+80~245).  We find, however, that 
the low-$\alpha$ stars do not share the same behaviour in the 
abundances of Fe-peak elements.  BD+80~245 exhibits slightly low 
Fe-peak abundances with respect to iron (0.2 -- 0.3~dex), whereas 
G4-36 and CS22966-043 are significantly enhanced (up to 0.5 -- 
0.7~dex) relative to the scaled solar metallicity.  
Table~\ref{table1} displays the abundances we derived for our
stars, as well as the mean of the halo in the metallicity range
of --1.75 $<$ [Fe/H] $<$ --2.25.  As can be seen in this table,
as well as in Figure~\ref{figure1}, the elemental abundances of 
our low-$\alpha$ stars are very dissimilar to those found in the 
general halo population.  Further details of the observational
material and abundance analysis are described in Ivans et al 
(2003).

 \begin{table}
  \caption{Abundance Results and Comparison to Halo Field Means}
    \begin{tabular}{lllll}
     \hline \hline
{Ratio}      & {Halo Field} & {BD+80~245} & {G4-36} & {CS22966-043} \\
     \hline
\feih&                      & --2.09 (0.11) & --1.93 (0.11) & --1.91 (0.15) \\
\feiih&                     & --2.04 (0.11) & --1.95 (0.11) & --1.91 (0.16) \\
     \hline
\xfe&                       &             &         &               \\
     \hline
Na  &  +0.03 ($\sigma$=0.38)& --0.41 (0.05) & --0.28 (0.08) & --0.64 (0.16) \\
Mg  &  +0.37 ($\sigma$=0.13)& --0.22 (0.11) & --0.19 (0.14) & --0.65 (0.15) \\
Al  & --0.10 ($\sigma$=0.45)& --1.33 (0.15) & --1.44 (0.15) & $<$--0.9 \\
Si  &  +0.42 ($\sigma$=0.15)& --0.11 (0.15) & --0.26 (0.15) & --0.97   \\
Ca  &  +0.31 ($\sigma$=0.13)& --0.18 (0.13) & --0.21 (0.11) & --0.24   \\
Sc  &  +0.11 ($\sigma$=0.26)& --0.42 (0.10) & --0.76 (0.05) & --0.76 (0.05) \\
Ti  &  +0.26 ($\sigma$=0.16)& --0.30 (0.05) &  +0.54 (0.07) &  +0.60 (0.10) \\
V   & --0.06 ($\sigma$=0.23)& --0.39 (0.10) &  +0.32 (0.05) & $<$+1.2 \\
Cr  & --0.06 ($\sigma$=0.10)&  +0.01 (0.06) &  +0.41 (0.06) &  +0.34 (0.06) \\
Mn  & --0.28 ($\sigma$=0.16)& --0.26 (0.08) &  +0.37 (0.09) &  +0.41 (0.12) \\
Co  & --0.27 ($\sigma$=0.39)& --0.18 (0.10) &  +0.33 (0.20) & $<$+0.5 \\
Ni  &  +0.00 ($\sigma$=0.14)& --0.09 (0.09) &  +0.48 (0.11) &  +0.54 (0.15) \\
Cu  & --0.63 ($\sigma$=0.15)& $<$--1.0      & $<$--0.7      & $<$+1.5 \\
Zn  &  +0.08 ($\sigma$=0.08)& --0.42 (0.15) &  +1.00 (0.05) &  +1.09 (0.15) \\
Ga  & $\cdots$(a)           & --0.30 (0.10) &  +0.58 (0.10) &  +1.75 (0.20) \\
Sr  &  +0.17 ($\sigma$=0.43)& --0.85 (0.10) & --0.65 (0.26) & --0.99 (0.05) \\
Y   & --0.13 ($\sigma$=0.20)& $<$--1.2      & $<$--0.7      & $<$+0.6 \\
Ba  & --0.01 ($\sigma$=0.32)& --1.89 (0.15) & --0.72 (0.13) & $<$--0.6 \\
La  &  +0.07 ($\sigma$=0.30)& --0.82 (0.15) & $<$+1.4       & $<$+3.4 \\
Eu  &  +0.41 ($\sigma$=0.26)& --0.64 (0.18) & $<$+0.4       & $<$+1.2 \\
     \hline \hline
    \end{tabular}

\hspace{8pt} Additional Note:\\
(a) Ga abundances for metal-poor stars do not exist in the literature.
  \label{table1}
\end{table}

\section{Comparisons to Supernova Model Yields}

The iron contribution from SNe~Ia yields to Galactic evolution is 
observed in the trend of the lowering of the [$\alpha$/Fe]-ratio at
[Fe/H] $\simeq$ --1 (Tinsley 1979; Pagel \& Tautvaisiene 1995, their
Figure 3).  In our study, we attempted to match the abundances of our 
unusual low-$\alpha$ stars to various yields from supernova models in 
the literature by making the assumption that the total abundance of 
some element ($X$) observed in these stars is the result of a mix of
SNe~Ia and SNe~II contributions:

\begin{equation}
{\cal R}_{obs} \equiv
\frac{M_{\star}(X)}{M_{\star}(Fe)} = 
 \frac{N_{Ia}M_{Ia}(X) + N_{II}M_{II}(X)}
  {N_{Ia}M_{Ia}(Fe) + N_{II}M_{II}(Fe)},\label{e-2}
\end{equation}

\noindent from which we can derive the following:

\begin{equation}
\frac{N_{Ia}}{N_{II}} = 
 \frac{M_{II}(X) - {\cal R}_{obs}~M_{II}(Fe)}
 {{\cal R}_{obs}~M_{Ia}(Fe) - M_{Ia}(X)} \label{e-3}
\end{equation}

\noindent where ${\cal R}_{obs} \equiv M_{\star}(X)/M_{\star}(Fe)$ 
denotes the ratio of the mass of element ($X$) to the mass of iron in 
our star, $M_{Ia}(X)$ and $M_{II}(X)$ denote the mass of element ($X$) 
ejected from SNe~Ia and SNe~II, and $N_{Ia}/N_{II}$ represents the 
ratio of the number of SNe~Ia to SNe~II events that fit the 
observations and the synthesized mass of elements ($X$) and iron from 
the model yields.  

We employed yields from models representing SNe~II from massive stars 
in the range of 10--50~M$_{\odot}$ and those representing SNe~Ia 
resulting from thermonuclear explosions of electron-degenerate cores 
such as those found in white dwarfs in close binary systems.  We used 
the yield compilation of Iwamoto et al (1999; their Table 3), which 
includes SNe~II yields based on calculations by Nomoto \& Hashimoto 
(1988), Hashimoto et al (1989, 1996) and Thielemann et al (1996), and 
SNe~Ia yields for seven models based on calculations by the Nomoto et 
al (1984), Thielemann et al (1986), Hix \& Thielemann (1996).  We also 
expanded the SNe~Ia model yields we employed to include a suite of 45 
models produced in calculations made by H\"oflich et al (1995, 1998, 
2002), H\"oflich \& Khokhlov (1996), and Dom\'inguez et al (2001).

We first tested the results of the model yields against the observed 
solar abundance ratios.  Using solar abundances of O, Na, Mg, Al, and 
Si from the compilation of Anders \& Grevesse (1989), and the SNe~Ia
yields tabulated in Table~3 of Iwamoto et al, we derived a value of 
$(N_{Ia}/N_{II})_{\odot}$ = 0.24 $\pm$ 0.07.  We then explored the 
differences that resulted from revising the Anders \& Grevesse 
recommendation of log~$\epsilon$(O) = 8.93 to 8.80, the O abundance 
recommended by Reetz (1999), 8.74 by Holweger (2001), and 8.69 by 
Allende Prieto et al (2001).  These O abundance revisions resulted in 
raising both the $(N_{Ia}/N_{II})_{\odot}$, and associated $\sigma$ to 
0.28 $\pm$ 0.07, 0.29 $\pm$ 0.09, and 0.31 $\pm$ 0.11, respectively.
Since we have no way to determine whether the increase in scatter
resulting from the revised O abundances are due to errors in the
abundances or to errors in the model yields, we dropped [O/Fe] from 
our $(N_{Ia}/N_{II})$ determinations, recognizing that the evidence 
for this decision is not entirely compelling.  The supernovae ratio 
we derive from [(Na,Mg,Al,Si)/Fe]$_{\odot}$ using this set of SNe~Ia 
model yields is $(N_{Ia}/N_{II})$ = 0.25 $\pm$ 0.06.

For other elements, the simple prescription fails.  As seen in 
previous studies (see eg.\ Timmes et al 1995, and references 
therein), the observations of the abundances of N, Sc, Ti, V, Co, 
Ni, Cu, and Zn, among others, show poor agreement with the
predictions by supernova models.  A better understanding of 
supernovae contributions to nucleosynthesis is required in order
to explain why no combination of supernova models is able to 
satisfactorily reproduce the observations of the abundances of 
these elements.  

As an additional ``sanity check'', we also derived the ratio of
SNe~Ia to SNe~II required to explain the observed abundances of 
CS22892-052, a metal-poor ([Fe/H] $\simeq$ --3.1) $r$-process-rich 
star believed to have witnessed one (or only a few) previous 
supernova(e) event(s) of Type II only (see eg.\ Sneden et al 2003, 
and references therein).  The SNe~Ia to SNe~II ratio we derive from 
the published ratios of [(Na,Mg,Si)/Fe] is 
$(N_{Ia}/N_{II})_{CS22892-052}$ = 0.01 $\pm$ 0.11, a sensible 
result.  In our application of the method to stars other than the 
Sun, we found that the results for Al were not well-matched to 
those from Na and the $\alpha$-elements.  Discussion in the 
literature has noted that the abundance of light-odd-Z neutron-rich 
elements are sensitive to the neutron excess and errors in the 
models can be large.  Rederiving the supernovae ratio for the Sun, 
without including the Al abundance, we find 
$(N_{Ia}/N_{II})_{\odot}$ = 0.22 $\pm$ 0.05.  Employing the 
expanded suite of models by the H\"oflich group, we derive similar 
results: $(N_{Ia}/N_{II})_{\odot}$ = 0.18 $\pm$ 0.01.  

The ratio of contributing supernovae types we derive for the Sun 
are in good agreement with literature values of Galactic ratios of 
SNe~Ia to SNe~II events, as shown in Table~\ref{table2}.  For the 
average abundances listed in Table~\ref{table1}, we derive 
$(N_{Ia}/N_{II})_{Halo}$ = 0.09 $\pm$ 0.11 (employing yields 
tabulated in Iwamoto et al) or $\pm$ 0.10 (employing the H\"oflich 
group SNe~Ia yields).  However, the ratios we derive for the 
low-$\alpha$ stars of this study (as well as some of the stars
discussed in \S\ref{intro}), require significantly greater amounts 
of SNe~Ia material.

 \begin{table}
  \caption{$(N_{Ia}/N_{II})$ Estimates}
    \begin{tabular}{clcl}
     \hline \hline
{Star} & {$(N_{Ia}/N_{II})$} & {[$X$/Fe]} & Source \\
     \hline
Sun    & 0.25 $\pm$ 0.06 & Na, Mg, Al, Si & (1) \\
       & 0.22 $\pm$ 0.05 &     Na, Mg, Si & (1) \\
       & 0.18 $\pm$ 0.01 &     Na, Mg, Si & (2) \\
       & 0.12 -- 0.36    & $\cdots$       & (3) \\
       & 0.15            &                & (4) \\
       & 0.187 -- 0.257  &                & (5) \\
CS22892-052 & 0.01 $\pm$ 0.11 & Na, Mg, Si & (1) \\
            & 0.01 $\pm$ 0.10 & Na, Mg, Si & (2) \\
Halo [Fe/H] $\simeq$ --2
       & 0.09 $\pm$ 0.11 & Na, Mg         & (1) \\
       & 0.09 $\pm$ 0.10 & Na, Mg         & (2) \\
BD+80~245 & 0.58 $\pm$ 0.21 & Na, Mg, Si & (1) \\
          & 0.47 $\pm$ 0.21 & Na, Mg, Si & (2) \\
G4-36     & 0.43 $\pm$ 0.10 & Na, Mg     & (1) \\
	  & 0.45 $\pm$ 0.12 & Na, Mg, Si & (2) \\
CS22966-043 & 1.29 $\pm$ 0.08 & Na, Mg, Si & (1) \\
            & 1.22 $\pm$ 0.01 & Na, Mg, Si & (2) \\
NS97 Low-$\alpha$ & 0.23 $\pm$ 0.11 & Na, Mg, Si & (1) \\
                  & 0.20 $\pm$ 0.10 & Na, Mg, Si & (2) \\
NS97 Other Halo   & 0.08 $\pm$ 0.03 & Na, Mg, Si & (1) \\
                  & 0.07 $\pm$ 0.03 & Na, Mg, Si & (2) \\
Pal 12    & 0.38 $\pm$ 0.22 & Na, Mg, Si & (1) \\
          & 0.29 $\pm$ 0.18 & Na, Mg, Si & (2) \\
Rup 106   & 0.39 $\pm$ 0.32 & Na, Mg, Si & (1) \\
          & 0.34 $\pm$ 0.31 & Na, Mg, Si & (2) \\
HD134439/40 & 0.41 $\pm$ 0.18 & Na, Mg, Si & (1) \\
            & 0.34 $\pm$ 0.19 & Na, Mg, Si & (2) \\
     \hline \hline
    \end{tabular}

\hspace{8pt} Table Notes and References:\\
(1) Derived from fits of stellar abundances to published SNe~II and SNe~Ia yields (as tabulated in Iwamoto et al 1999; their Table 3).\\
(2) Derived from fits of stellar abundances to published SNe~II (as tabulated in Iwamoto et al 1999; their Table 3) and SNe~Ia yields (from the H\"oflich group).\\
(3) van den Bergh \& Tammann (1991)\\
(4) Tsujimoto et al (1995) \\
(5) Iwamoto et al (1995)
  \label{table2}
\end{table}

\subsection{Results from SNe~Ib, SNe~Ic, and High Energy SNe~II (``Hypernovae'') Yields}

Employing the yields of hypernovae models of varying mass and energy
developed by Umeda \& Nomoto (2002) in place of the SNe~II model 
calculations tabulated by Iwamoto et al, we rederived the 
$(N_{Ia}/N_{II})$ ratios for our low-$\alpha$ stars.  In no instance does 
the variation in derived $(N_{Ia}/N_{II})$ improve over those of standard 
SNe~II.  We also rederived the $(N_{Ia}/N_{II})$ ratios for our 
low-$\alpha$ stars employing yields of the full range of SNe~Ib and 
SNe~Ic models calculated by Woosley et al (1995).  We attempted to employ 
their yields to our abundance ratios both by treating the SNe~Ib and
SNe~Ic models as alternatives to the SNe~II yields, and by treating them 
as alternatives to the SNe Ia yields.  In neither case could we obtain 
satisfactory agreement among the abundances for any of our stars.  We
find that the unusual abundances of our stellar trio cannot be explained
by the hypernova models of Umeda \& Nomoto or by the SNe~Ib or SNe~Ic 
models of Woosley et al.

\section{Alternative Explanations?}

We explored the possibility that the source of the abundance anomalies
in our stars may be the result of something other than SNe~Ia-enhanced
material.  The details are presented in Ivans et al (2003) but, in
summary, we find we cannot explain the abundances by invoking binarity
and mass transfer; formation from AGB-enriched material; or chemically
stratified atmospheres.  The supernovae ratios we derived do not allow
us to distinguish between the possibility that our stars formed from 
material enriched in SNe~Ia yields and the possibility that our stars
were born from much less chemically enriched material at earlier times
and nearby SNe~Ia events occurring at later times deposited Fe nuclei 
on the surfaces of our stars.  However, the ratios of the abundances
of the neutron-capture elements suggest that the latter possibility is 
less likely.

{\it While it is possible that the atmospheres do not represent 
the interstellar material out of which they were born, and instead 
exhibit the effects of pollution from some other source, no potential 
contributor has been observed that displays abundances similar to those 
we derive.}

\section{Summary \& Conclusions}

Evidence of chemical substructure in the Galactic halo is seen
in the elemental abundances of BD+80~245, G4-36, and CS22966-043.
These stars were all serendipitously discovered as having 
low $\alpha$-element and neutron-capture element abundances in
the context of previous studies.  In our study, we confirm the
earlier reports and expand the number of elements for which
abundances have been derived in these stars.  We find that the
mean abundance ratio for [(Mg,Si,Ca)/Fe] is $\sim$ 0.7~dex 
below the mean of halo stars of comparable metallicities and
that the mean abundance ratio for [(Sr,Ba)/Fe] is $\sim$ a 
full dex below the halo mean. 

Our low-$\alpha$ stars do not display uniform abundances of the
Fe-peak nuclei with respect to the abundance of iron.  BD+80~245
exhibits diminishments whereas G4-36 and CS22966-043 present 
mild to pronounced over-abundances of these elements.  The latter
two stars require such large Ga enhancements to synthesize the
observed spectra that it is possible, for the first time, to
report Ga abundances in any metal-poor star.

We explore the idea that the elemental abundance ratios in these
stars can be explained by iron enhancements from SNe~Ia 
contributions.  We derive 0.18 $\pm$ 0.01 $<$ $(N_{Ia}/N_{II})$ 
$<$ 0.25 $\pm$ 0.06 for the Sun, which is in good agreement with
values found in the literature for Galactic ratios of supernovae
Type II to Type Ia events.  The results for our low-$\alpha$
trio confirm the idea that the low $\alpha$- and low 
neutron-capture element abundances can be explained by a larger 
contribution of SNe~Ia yields than went into the average halo
star: 0.6 $\le$ $<$ $(N_{Ia}/N_{II})$ $<$ 1.3.  Thus, given the
metallicities of our stars, it would appear that BD+80~245, G4-36, 
and CS22966-043 may have formed from the material polluted by the 
earliest of SNe~Ia events.  However, our derivations were limited 
to a small subset of the observed abundances.  The unusual 
abundance ratios of Ti, Cr, Mn, Ni, and Zn (among other elements) 
cannot be modelled using existing supernova model yields.  
Additional questions which we would like to explore are the 
determination of the SNe~Ia progenitors, and whether the 
progenitors or scenarios have changed over Galactic timescales.
To our knowledge, an identification or determination of SNe~Ia
progenitors based on the observed chemical abundances of 
metal-poor stars does not exist in the literature.  The three
low-$\alpha$ stars represent a non-negligible part of early
Galactic nucleosynthesis, yet are not explained by current 
Galactic evolution models.  

\section{Acknowledgments}
Research by III is currently supported by NASA through Hubble 
Fellowship grant HST-HF-01151.01-A from the Space Telescope Science 
Institute, which is operated by the Association of Universities for 
Research in Astronomy, Inc., under NASA contract 5-26555.  I also
thank the organizers of the Carnegie Observatories Symposium on the 
Origin and Evolution of the Elements for making this meeting such 
an interesting and fruitful one.

\end{document}